\documentclass[proof]{WileyASNA-v1}

\articletype{Article Type}%

\received{}
\revised{}
\accepted{}

\begin{document}

\title{Hot Jupiters accreting onto their parent stars: effects on the stellar activity}

\author[1]{Salvatore Colombo*}
\author[1]{Ignazio Pillitteri}
\author[1]{Salvatore Orlando}
\author[1]{Giuseppina Micela}

\authormark{Colombo, S. \textsc{et al}}

\address[1]{\orgdiv{INAF}, \orgname{Osservatorio Astronomico di Palermo}, \orgaddress{\state{1 Piazza del Parlamento, Palermo, 90134}, \country{Italy}}}

\corres{*Salvatore Colombo. \email{salvatore.colombo@inaf.it}}

\presentaddress{}

\abstract{Hot Jupiters (HJs) are massive gaseous planets orbiting close to their host stars. 
Due to their physical characteristics and proximity to the central star, HJs are the natural laboratories to study the process of Stellar Planet Interaction (SPI). Phenomena related to SPI may include the inflation and evaporation of planetary atmospheres, the formation of cometary tails and bow shocks and magnetospheric interaction between the magnetic field of the planet and that of the star. Several works suggest that some systems show enhanced stellar activity in phase with the planetary rotation period. 
In this work, we use a 3D magneto-hydrodynamic model that describes a system composed by a star and a HJ and that includes the corresponding planetary and stellar winds. The aim is to investigate whether the material evaporating from the planet interacts with the stellar extended corona, and generates observable features. 

Our simulation shows that, in some conditions, the planetary wind expands and propagates mainly along the planetary orbit. Moreover, part of the planetary wind collides with the stellar wind and a fraction of the planet’s outflow is funneled by the stellar magnetic field and hits the stellar surface. In both events the material is heated  up to temperatures of a few MK by a shock. These phenomena could manifest in the form of enhanced stellar activity at some orbital phases of the planet. 
}




\maketitle


\section{Introduction}\label{intro}
Hot Jupiters (HJs) are massive (1 to 10 Jupiter masses) gaseous planets orbiting very close (semi major axis $\le 0.1 AU$) to their host stars. Due to their mass and proximity to the central star, HJs are natural laboratories to study the stellar-planet interaction (SPI). 
The atmospheres of HJs are heated up by the radiation arising from the central star \citep{Burrows2000,OwenJackson2012,Buzasi2013,Lanza2013}.
In some cases, the heated planetary gas has enough thermal energy to escape from the gravitational field of the planet \citep{Lammer2003}. This phenomena is called photoevaporation. Evidence of observations of these outflows have been reported for example for the systems HD 189733b and HD 209458b \citep{Vidal-Madjar2004,Ehrenreich2008,Lecavelier2010,Lecavelier2012,Bourrier2013}. 
The stellar wind is the medium where the planetary outflow expands and interacts. This interaction could produce observable signatures that may results in an enhanced stellar activity.

The system HD 189733 represents one of the best targets to study SPI because of its proximity, strong activity, and the presence of a transiting planet, which allows the observation of planetary photoevaporation.
X-ray observations of this system suggest a correlation between the planetary transit and the stellar activity \citep{Pillitteri2010, Pillitteri2011, Pillitteri2014}. 
An interpretation about the origin of the X-ray emission is proposed by \cite{Pillitteri2015}. 
They suggest that part of the evaporating material from the planet is intercepted by the stellar gravitational field and accretes  onto the star forming an accretion column. The impact of the supersonic accretion column with the stellar surface generates a hotspot which is responsible for the enhanced UV and X-ray emission. 

This idea follows from the work by \cite{Matsakos2015}, where a SPI magneto-hydrodynamic (MHD) model was developed to describe the various effects of SPI expected in an idealized star-planet system.
Depending on the physical conditions of the system different kinds of interaction are predicted. 
The planetary outflow may be forced by a strong stellar wind in a cometary tail which follows the planet in its orbit \citep{Matsakos2015} and produces observable signatures during the planetary transit \citep{Mura2011,Kulow2014}. 
Conversely, the planetary wind may produce a bow shock during the expansion in the stellar wind \citep{Vidotto2010,Vidotto2011a,Vidotto2011b,Llama2011,Llama2013}. 
Then, in some cases, it could be catched by the gravitational field of the central star. When this occurs the planetary wind is forced to spiral down into the star. The stellar magnetic field funnels the falling gas onto an steady accretion column-like structure. The impact of this accretion flow with the stellar surface produces a hotspot that may generate an observable excess of UV and X-ray radiation \citep{Lanza2013,Matsakos2015,Pillitteri2015}. The hotspot orbits with the planetary period, but shifted in phase of $\approx{90}^{\circ}$.
Although the formation of the bow shock in the case of HJs is still debated, its existence has been claimed by several authors \citep[see for example:][]{Llama2011,Llama2013,Murray-Clay2009,Matsakos2015,Vidotto2011a,Cohen2011,Bisikalo2013}.

Despite the observational evidence \citep{Pillitteri2015} supported by MHD modeling \citep{Matsakos2015}, there are other explanations for the X-ray activity in HD 189733. In fact, \cite{Route2019} claim that there are no statistical evidence for a bright hotspot synchronized to the planetary orbital period. 
Using Kolmogorov-Smirnov and Lomb-Scargle periodogram analyses, they did not find evidence for persistent hotspots that have locations synchronized to the planetary orbital period.
They suggest that the bright regions that persist for a few rotational periods, are entirely consistent with the normal evolution of active regions on stars.

In this work we focus on HD 189733 and we investigate whether and in which configurations/conditions the planetary wind is accreting onto the star as predicted by \citep{Pillitteri2015}. 
To address this question we developed an MHD model that describe the star-planet system. The model includes the gravity and the magnetic field from both the star and the planet. We assume typical values, for this kind of system, for both stellar and planetary winds.

The paper is structured as follows: In Sect. \ref{model} we describe the MHD model used in this work, in Sect. \ref{results} we discuss the results from the simulation. Finally, in Sect. \ref{conclusions} we draw our conclusions.

\section{Model}\label{model}
Star-planet systems with evaporating HJ can be fully described making use of hydrodynamic or magnetohydrodynamic (MHD) simulations, as discussed in \cite{Matsakos2015} (see also reference therein).  In this work, we adopted the MHD described in \cite{Matsakos2015}, but tuned on the system HD 189733 \citep[see also][]{Colombo2021}.

HD 189733 is a K1.5V type star with mass $M_{s}=0.805$M$_\sun$ radius $R_{s}=0.76R_{\sun}$, and a rotational period of $\approx11.95$d \citep{Henry2008} at a distance of $\approx19.3$ pc from the Earth. It is part of a binary system with the companion star orbiting at a mean distance of $D_s\sim 220$ AU. HD 189733A hosts a HJ with mass $M_{p}=1.13$M$_J$ that orbits at a distance of $D_p \sim 0.031$AU with a period of $2.219$ days \citep{Bouchy2005}.   

The model setup adopts Cartesian coordinates ($x$,$y$,$z$). Since $M_{s}\gg M_{p}$ we consider the center of the star as center of mass of the entire system. The planet orbits around the star in the x-y plane. 

\subsection{Equations}
The model solves the MHD equations, 
\begin{gather}
	\frac{\partial}{\partial t}\rho + \nabla \cdot  \rho\vec{v} =0\\	
	\frac{\partial}{\partial t}\rho\vec{v}+\nabla \cdot (\rho\vec{v}\vec{v} - \vec{B}\vec{B}+ \vec{I}p_t) = \rho\vec{g}\\		
    \frac{\partial}{\partial t}\rho E+\nabla\cdot[(\rho E+p_t)\vec{v}-\vec{B}(\vec{v}\cdot\vec{B})] = \rho \vec{v}\cdot(\vec{g}+\vec{F}_{ext}) \label{RL} \\
	\frac{\partial}{\partial t}\vec{B}+\nabla\cdot(\vec{v}\vec{B}-\vec{B}\vec{v})=0 \\
	\frac{\partial\rho\sigma}{\partial t} + \nabla \cdot (\rho\sigma\vec{v}) = 0
\label{eq_Q}
\end{gather}
\noindent
where: 
\begin{equation}
p_t= P+\frac{\vec{B} \cdot \vec{B}}{2} , \qquad E = \epsilon + \frac {\vec{v} \cdot \vec{v}}{2}+\frac{\vec{B} \cdot \vec{B}}{2\rho}
\end{equation}
where, $\rho$ is the plasma density, $\vec{v}$ the plasma velocity, $\vec{I}$ the identity matrix, $p_t$ the total plasma pressure, $\vec{B}$ the magnetic field, $\vec{g}$ the gravity acceleration vector, $\vec{F}_{ext}$ is a inertial force that appears in our non-inertial rotating frame. $\vec{F}_{ext}=\vec{F}_{Coriolis}+\vec{F}_ {centrifugal}$ has a Coriolis and centrifugal components given by: $\vec{F}_{Coriolis} = 2(\Omega_{fr}\times \vec{v})$ and $\vec{F}_{centrifugal}=-\left[\Omega_{fr}\times(\Omega_{fr}\times \vec{R})\right]$. 
$E$ is the total plasma energy density, $\epsilon$ the thermal energy density and $\sigma= p/\rho^\gamma$ the plasma entropy.
We use the ideal gas law $P = (\gamma -1)/\rho\epsilon$, where the polytropic index $\gamma = 3/2$. 

The calculation is perfomed using the PLUTO code, a modular Godunov-type code for astrophysical plasmas \citep{Mignone2012}. The code uses parallel computers using the Message Passage Interface (MPI) libraries. The MHD equations are solved using the MHD module available in PLUTO with the Harten-Lax-van Leer Riemann solver. The time evolution is solved using the Hancock method. To follow the magnetic field evolution and to maintain the solenoidal condition we use the divergence cleaning technique, an approach based on the generalized Lagrange multiplier (GLM) formulation \citep{Mignone2012}.

\subsection{Initial and Boundary conditions}
\begin{figure}
    \centering
    \includegraphics[scale = 0.2]{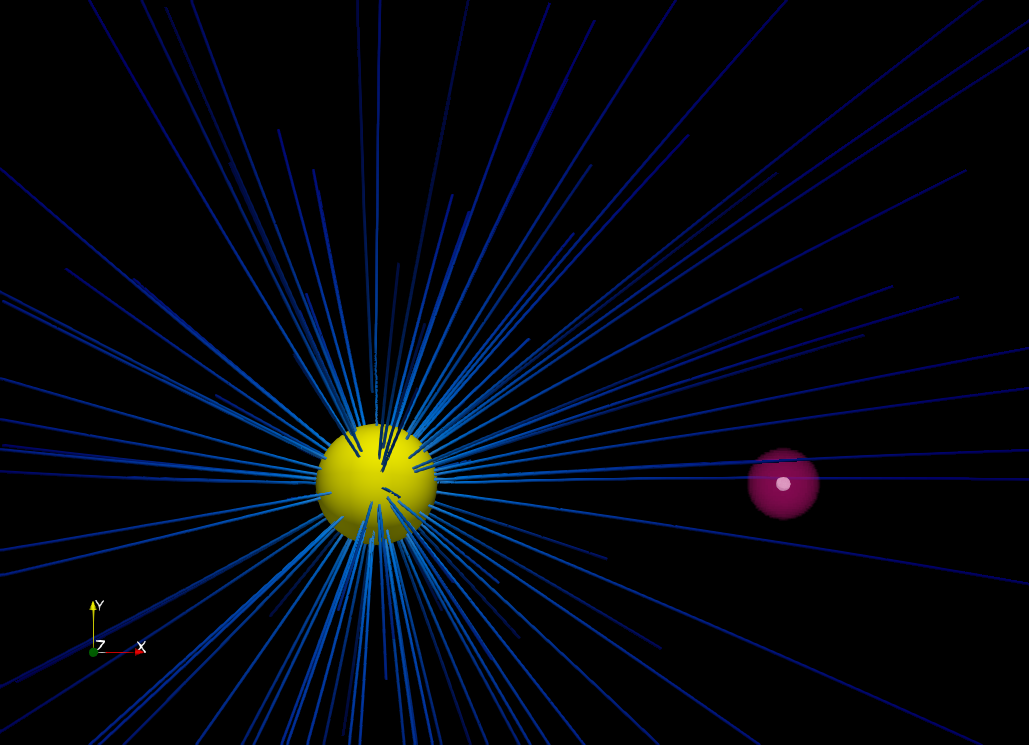}
    \caption{Density distribution at the beginning of the simulation. The initial planetary wind density is represented by using a volume rendering in red. The solid yellow sphere represents the surface of the star. The solid white tiny sphere, embedded in the planetary wind (in red), represents the surface of the planet. The blue lines are sampled magnetic field lines.}
    \label{ci}
\end{figure}

Initially, the star and the planet are surrounded by a low density static medium ($\rho = 10^{-19}$g cm$^{-3}$). The stellar magnetic field is described by the Parker spiral \citep{Parker1958}. The planetary magnetic field is assumed to have only a dipole component.
Fig. \ref{ci} shows the initial conditions for the simulation. 

At $t=0$, the stellar and planetary winds start to flow from the surface of the star and the planet respectively. We define two regions in proximity of the star and planet surfaces to generate the winds. The first is used to describe the planetary and stellar interiors. In this region $\rho$, $p$ and $\vec{B}$ are fixed to the values of the stellar and planetary winds but $\vec{v}$ is fixed equal to $0$.   
The second region between $R_{s/p}$ and $1.5 R_{s/p}$ generates the winds. There, $\rho$, $p$ and all the components of $\vec{v}$ are fixed to the values expected for the winds (see Table \ref{table:values}). To increase the code stability, we set the magnetic field in this region as free to evolve.
The chosen values of stellar and planetary winds are summarised in Table \ref{table:values}.
\begin{table}[h!]
\centering
\caption{Physical properties assumed for the stellar wind and the planetary outflow. For the stellar wind, we assumed a  density similar to that observed in the solar wind and a speed equal to the escape speed from the star; for the planet, we assumed an evaporation rate within the range of values discussed in the literature \citep{Foster2021} but slightly higher than that commonly accepted for HD 189733b \citep{Bourrier2013}.}
\label{table:values}
    \begin{tabular}{c|c|c}
                          & Star            & Planet \\
                            \hline
   Density (g cm$^{-3}$)  & $10^{-17}$      & $10^{-15}$  \\ 
   Temperature (K)        & $10^6$          & $10^4$              \\
   Velocity (cm s$^{-1}$) & $6\times10^{7}$ & $6.5 \times 10^{6}$ \\
   $\dot{M}$ (g s$^{-1}$) & $\approx 2\times 10^{13}$              & $\approx 6\times 10^{12}$ \\
   B (G) & 1& 0.1 \\
   Points on grid   &  $\approx 40$ & $\approx 10$ \\
   \end{tabular}
\end{table}

Outflow (zero-gradient) boundary conditions are set at all the external boundaries.  
The stellar and planetary interior are not involved in the calculation and they are prescribed as internal boundaries. To avoid numerical issues, we describe approximate equilibrium in these regions. Gravity is calculated as $g = \frac{4}{3}\pi G\rho R$, and pressure is given by $P_{int} = p+\frac{2}{3}\pi G \rho^2 (R_{sph} -R)$, where $R_{sph}$ is either the stellar or the planetary radius, accordingly.
The magnetic field is constant and uniform inside both the star and the planet. We approximate the star and the planet as rotating solid bodies that rotate around the axis parallel to the z-axis.

\subsection{Spatial Grid}
Our simulation requires a grid that adequately resolves the small scale structures of the planetary wind in a large domain. For this purpose, a static grid is too costly from the numerical point of view. Our strategy to reduce the numerical cost is twofold and follows the approach adopted by \cite{Matsakos2015}. As reference frame, the model adopts the corotating frame of the planet. In this way, the domain can be substantially reduced. In addition, we make use of the adaptive mesh refinement (AMR) module provided in PLUTO \citep{Mignone2012}. AMR allows us to increase the spatial resolution only where necessary, substantially reducing the numerical cost of the simulation. 

The domain is a Cartesian box where the point $(x,y,z)\equiv(0,0,0)$ is assumed to be the center of the star. The box extends from $-5$ to $15$  R$_{\sun}$ on the x-axis, from $-4$ to $16$ R$_{\sun}$  on the y-axis and from $-5$ to $5$ R$_{\sun}$  on the z-axis. We use a grid with 3 refinement levels.
The level 0 is composed of $128\times128\times64$ grid points. Each level is assumed to be a factor 2 more resolved than the previous one. In this way the 3rd refinement level is equivalent to a grid with $512\times512\times256$ points, corresponding to a resolution of $2.7 \times 10^4$km in each direction.
The grid is refined when the planetary gas fills more than 10\% of the cells.

In order to avoid an excess of refinement in regions which are not relevant for the dynamics, we imposed to refine only the cells made of a minimum of $10\%$ of planetary material.

\section{Results}\label{results}
In this section we analyse the planetary wind dynamics and, in particular, its interaction with the stellar magnetosphere and the possible effects on the resulting stellar activity. 
\subsection{Dynamics}

\begin{figure*}[!h]
    \centering
    \subfloat[a]{
        \includegraphics[scale = 0.15]{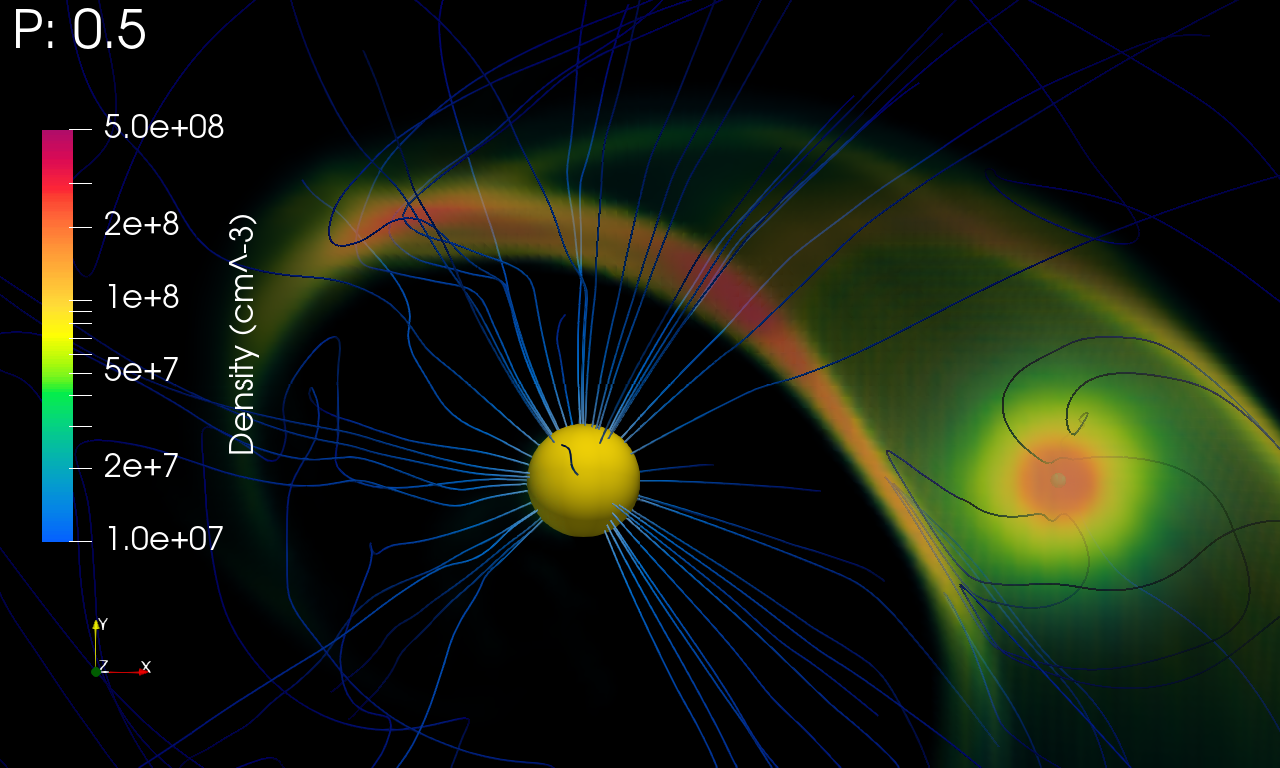}
        \label{ev_a}
    }
    \subfloat[b]{
        \includegraphics[scale = 0.15]{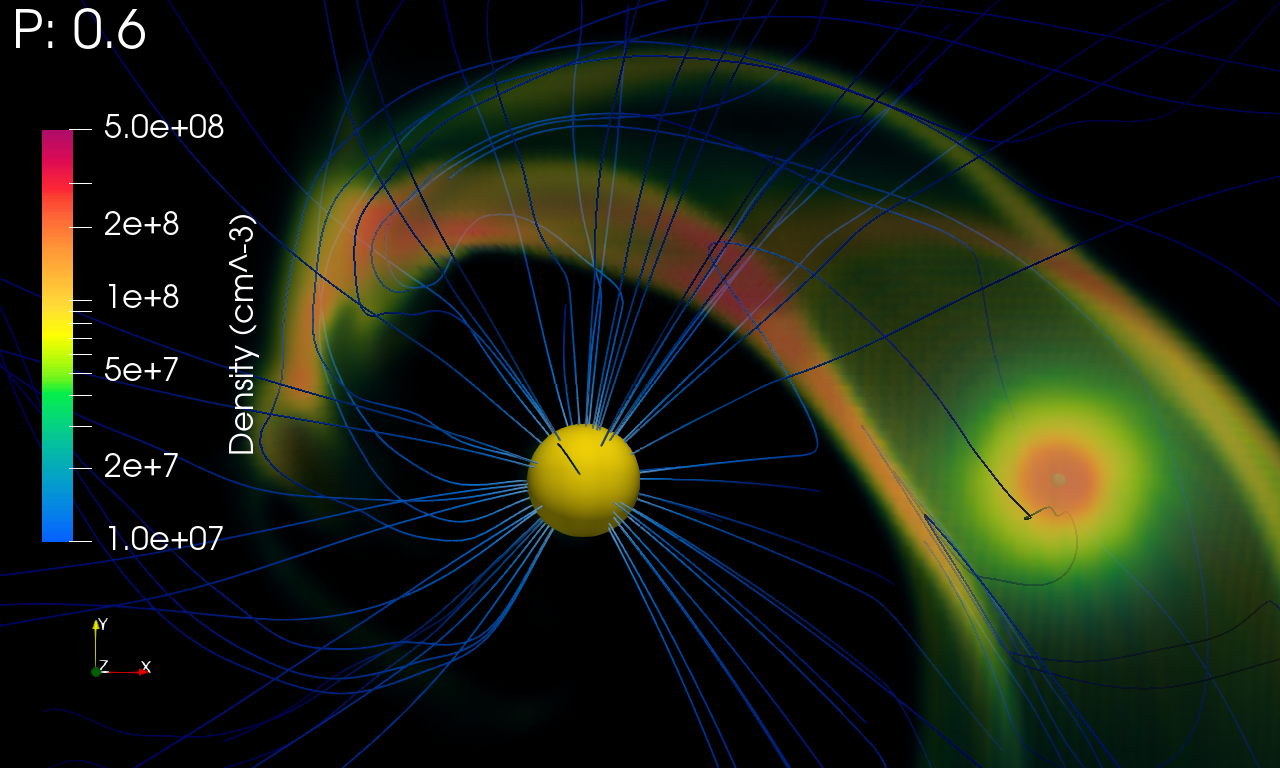}
        \label{ev_b}
    }
    \hspace{0mm}
    \subfloat[c]{
        \includegraphics[scale = 0.15]{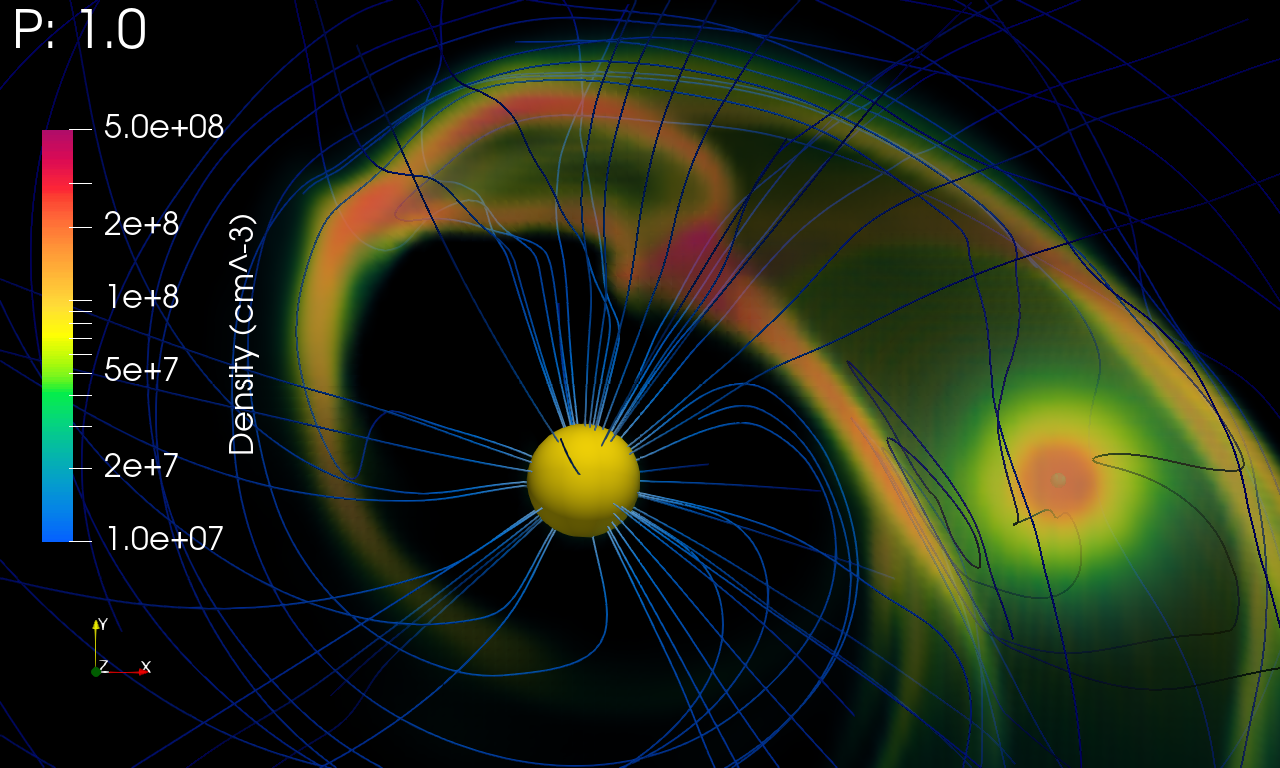}
        \label{ev_c}
    }
    \subfloat[d]{
        \includegraphics[scale = 0.15]{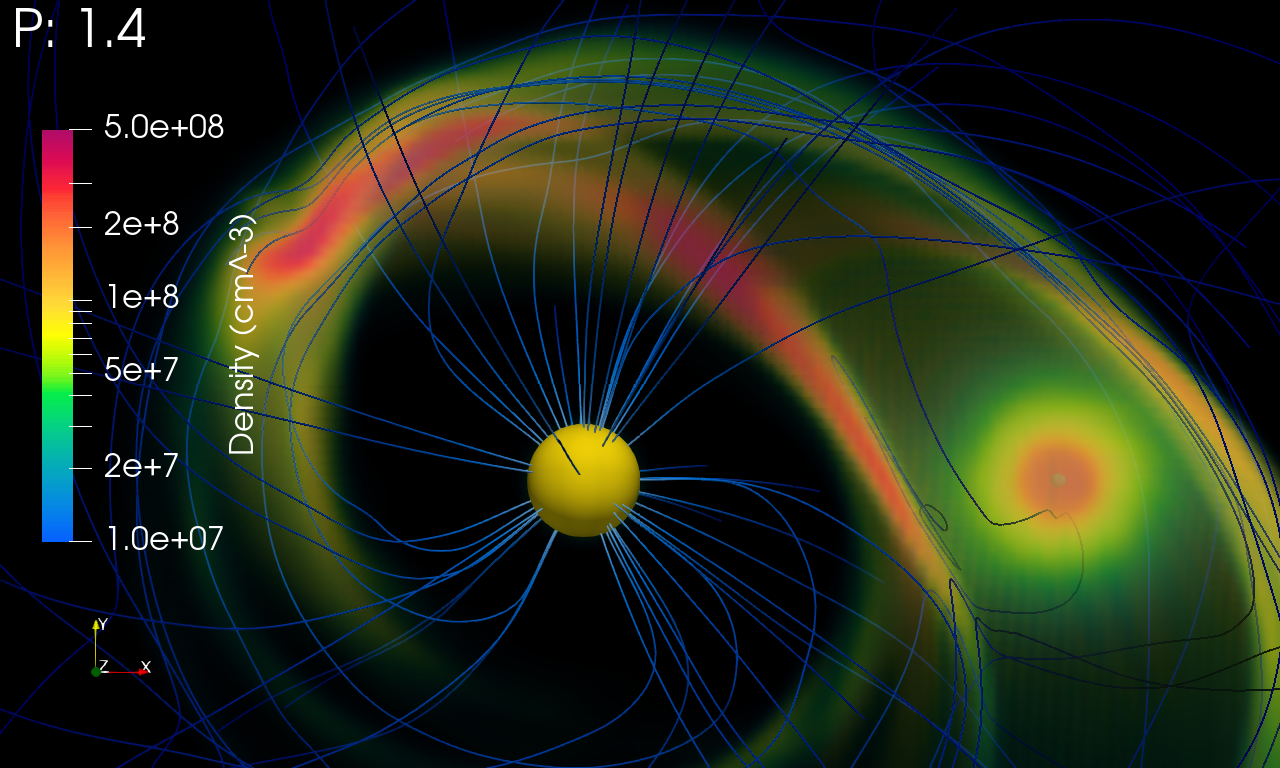}
        \label{ev_d}
    }
\caption{Evolution of the planetary wind during one orbit of the planet. The planetary wind is shown via volume rendering using a log colormap from blue to red. The yellow sphere in the middle represents the stellar surface (color not in scale). The blue tiny sphere represents the planetary surface (color not in scale). The blue lines represent the magnetic field lines. \label{ev} }
\end{figure*}

Fig. \ref{ev} shows the evolution of the planetary wind in different phases of the orbit.
Initially, the evolution is characterized by a transient phase, lasting approximately half period of the planet orbit, before reaching the regime condition. This transient is discarded from the analysis. Then the evolution is similar to the case called type III in \cite{Matsakos2015}. The entire dynamics can be divided into three different phases: expansion, accretion, disruption.

During the first phase, the planetary wind expands and propagates along the planetary orbit (Fig. \ref{ev_a}). The wind forms a large eye-shaped region around the planet (Fig. \ref{ev_a}). In this region the wind remains unperturbed during the whole simulation. On the outer region, the plasma dynamic is more complex. 
Most of the material, due to the planetary orbital motion, forms a cometary tail behind the planet. On the other side, the planetary wind expands ahead of the planet (Fig. \ref{ev_b}) 
 Due to its fast expansion and the collision to the stellar
wind, a shock front at temperature $T\approx10^6$ develops and
heats the planetary wind up to $T\approx10^5$K.
The shock regions are identified when two conditions are met: $\vec{\nabla} \cdot \vec{v} < 0 $ and $\Delta x \cdot \frac{\nabla p}{p} > \epsilon$, where $\epsilon = 5$ is a parameter that sets the shock strength. 

In this region the material originating from the planet and the stellar magnetic field strongly interact. The result of this interaction is a complex magnetic field configuration that may lead to magnetic reconnection phenomena \citep{Lanza2012}. 
The interaction with the stellar wind and the stellar magnetic field slows down the planetary material which is cached by the stellar gravity. During the falling the material is funneled in an accretion column that spirals around the star and eventually accretes onto the star (Fig. \ref{ev_c} accretion phase). 

The accretion onto the star is not stationary. The accretion column is attracted by the star due to the stellar gravity, but is also pushed away by the stellar wind. After $\approx 2\times 10^4$s, the disruption phase starts; the accretion stops and the stellar wind destroys the whole accretion column, pushing the material away from the system (Fig. \ref{ev_d}). 
Once the disruption phase terminates the system starts again to build up a new stream that is launched towards the star, thus starting another cycle in a quasi periodic regime. 

\subsection{Hotspots}

The stellar surface is perturbed by hotspots arising from the impact of the planetary wind onto the stellar surface. Before describing the hotspots evolution, it is worth to notice that the model does not include a description of the chromosphere and the confined corona of the star, both necessary to follow in details the accretion impacts. In fact, studying in detail the impacts of the planetary material onto the stellar surface is beyond the purpose of this work. Nevertheless, we can identify the regions on the stellar surface where we expect to observe the hotspots and describe their evolution.

\begin{figure}[!h]
\centering
    \subfloat[a]{
        \includegraphics[scale = 0.15]{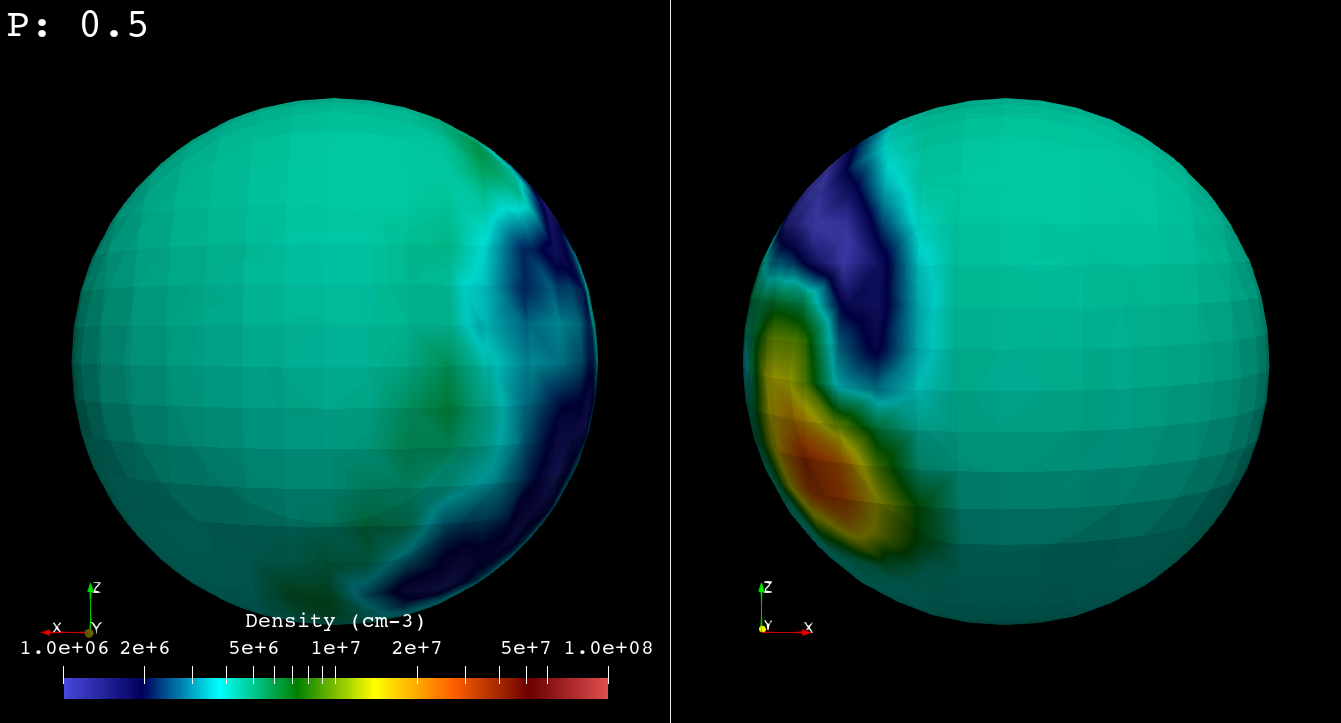}
        \label{hs_a}
    } \\
    \subfloat[b]{
        \includegraphics[scale = 0.15]{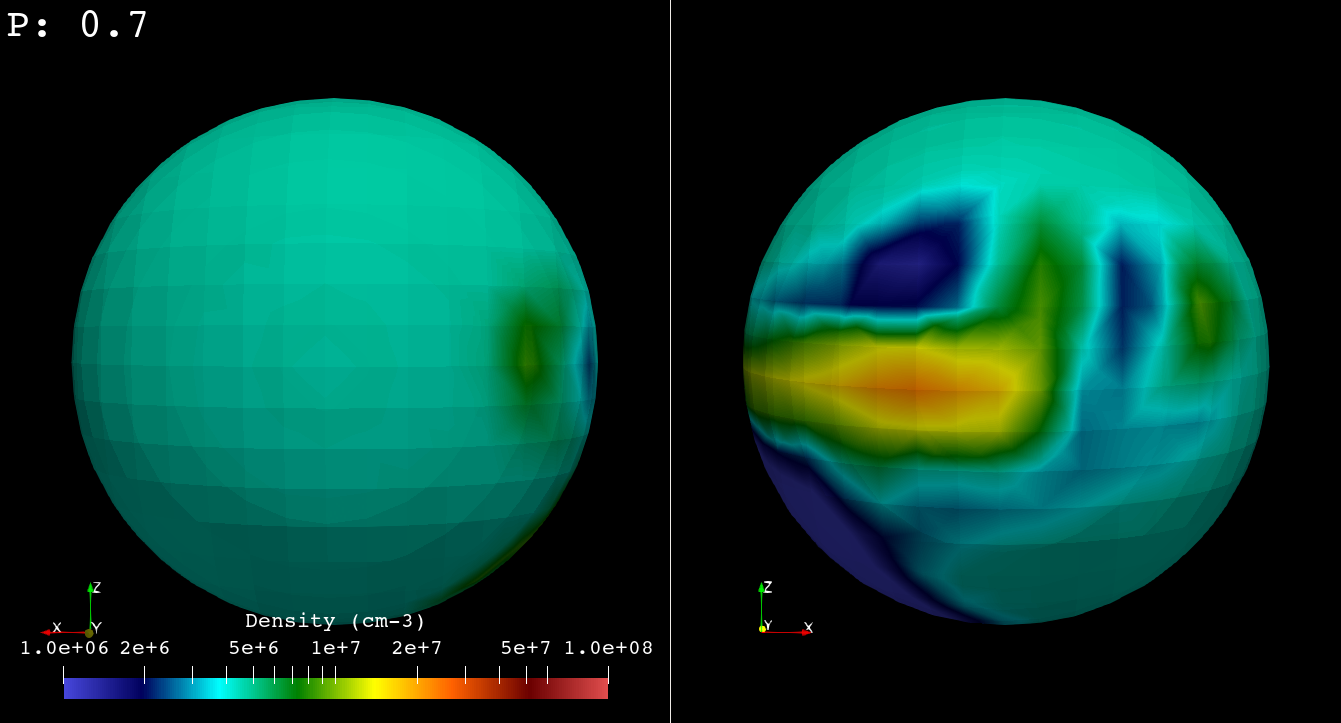}
        \label{hs_b}
    }\\
    \subfloat[c]{
        \includegraphics[scale = 0.15]{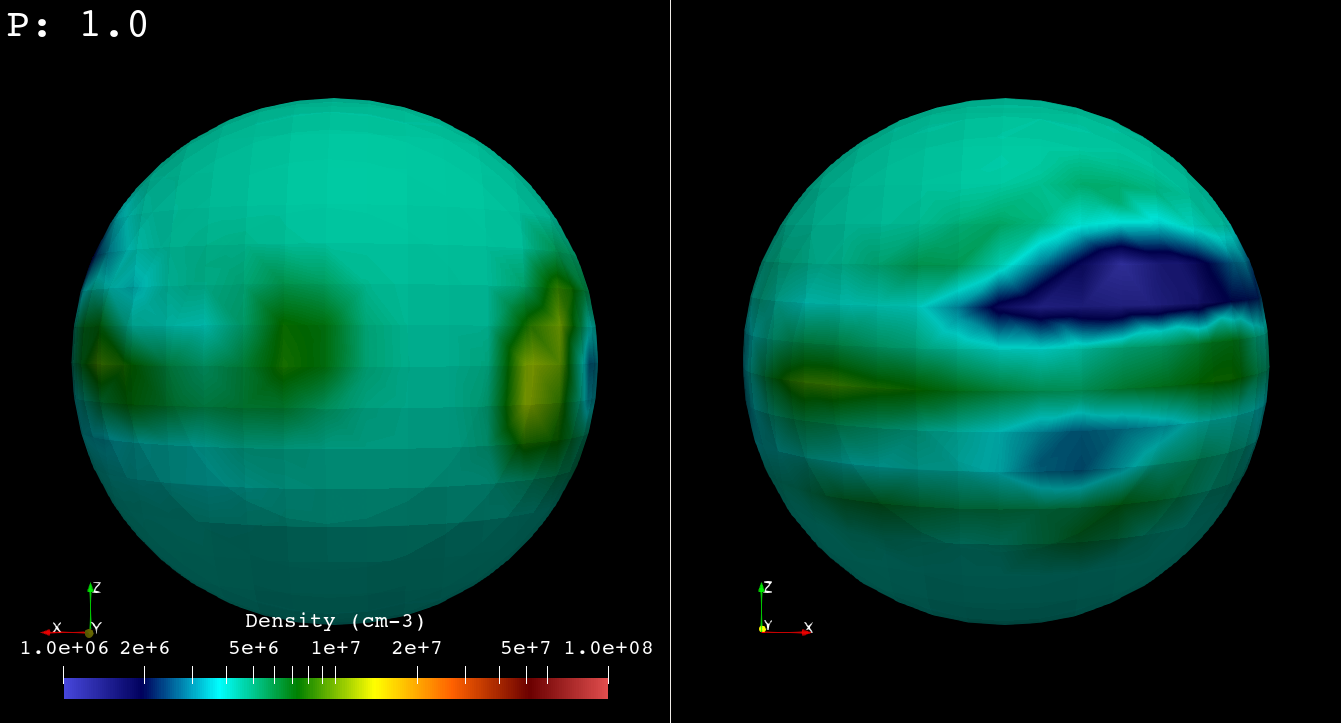}
        \label{hs_c}
    }
    \caption{ Logarithmic density maps for each face of the stellar surface (left and right panels) at three different times (from top to down panels). The star rotates from left to right in both panels. \label{hs}}
\end{figure}

Fig. \ref{hs} shows the evolution of the hotspots on the stellar surface predicted by the model.
These regions are expected to reflect the quasi periodic evolution observed for the planetary accretion column. 
Initially, only one faint hotspot is present on the stellar surface (Fig. \ref{hs_a}). The hotspot is located close to the stellar equatorial plane. The hotspot is denser in the core, with a maximum density $\approx 3\times10^7$cm$^{-3}$ (Fig. \ref{hs_a}). During the evolution, due to the stellar wind the accretion column is disrupted and the hotspot become fainter (Fig. \ref{hs_b}) and, then, at later time disappears (Fig. \ref{hs_c}). 

The density values in the accretion column are lower than values of a stellar corona. The velocity of the accretion column close to stellar boundary (at 1.5 $R_{s}$) is about $3\times 10^7$cm s$^{-1}$, so we expect impact speed slightly higher than this value. According to Eq. 8 from \cite{Sacco2010} the temperature of the post shock plasma due to the impact is $T_{ps} = 1.4\times10^{-9} u^2 \approx 10^6$K. 
Due to the low density of the accretion stream, the X-ray emission arising from the impact of the stream onto the stellar surface is negligible compared to the X-ray emission arising from the stellar corona. A higher wind density would be required to increase the X-ray emission from the hotspot. Since we already assumed a relatively high evaporation rate for the planet, we conclude that, at least in this particular case, the SPI can not produce detectable emission in the X-ray band.

\section{Conclusions}\label{conclusions}

In this paper, we investigate whether the planetary wind accretes onto the star in the system HD 189733 as hypothesized by \cite{Pillitteri2015}. To this end, we adopted a 3D MHD model analogous to that presented in \cite{Matsakos2015}. The model is tuned to the system HD 189733. Our results can be summarised as follow:
\begin{itemize}
    \item We find evidence that in the case of HD 189733, the planetary wind induced by photoevaporation expands along the orbit of the planet and forms a cometary tail behind the planet and a finger ahead of the planet that spirals around the star and, eventually, accretes onto the stellar surface. 
    \item The accreting planetary wind can produce few hotspots.
\end{itemize}

Our simulation is a starting point for more detailed works.
In a future project, we plan to include an exploration of the parameter space (Colombo et al. in prep.). We plan to explore the effects of different planetary wind density and the magnetic field strength. In fact, we expect that a denser planetary wind could lead to a denser accretion column. Moreover, the magnetic field intensity is expected to play a role in the dynamics of the planetary wind: a more intense magnetic field shield more the planet from the stellar wind and allows the planetary wind to expand more before interacting with the stellar wind. This may lead to an increase of the density and the stability of the accretion column. 

The synthesis of the emission in the X-ray and UV band is a further step that we plan to include in our work. The aim is to compare our model with the observation from \cite{Pillitteri2015} in order to confirm or not if the enhanced flaring activity observed in HD 189733 is a result of SPI. 

\section*{Acknowledgements}
We acknowledge support from ASI-INAF agreement  2021-5-HH.0 Partecipazione alla fase B2/C della missione ARIEL (Atmospheric Remote-Sensing Infrared Exoplanet Large-survey).
PLUTO is developed at the Turin Astronomical Observatory in collaboration with the Department of Physics of Turin University. We acknowledge the “Accordo Quadro INAF-CINECA (2017)” the CINECA AwardHP10CL7BPQ and the HPC facility (SCAN) of the INAF – Osservatorio Astronomico di Palermo, for the availability of high performance computing resources and support.

\bibliography{Wiley-ASNA}
\end{document}